\documentclass{article}
\usepackage{graphicx} 
\usepackage{multicol}
\usepackage{multirow}
\usepackage{adjustbox}
\usepackage[margin=2.5cm]{geometry}
\usepackage{placeins}
\usepackage{authblk}

\title{\textbf{Underestimation of lung regions on chest X-ray segmentation masks assessed by comparison with total lung volume evaluated on computed tomography}}
\author[1]{\small \textbf{Przemysław Bombiński \thanks{contact details: Przemysław Bombiński MD, PhD, e-mail: przemyslaw.bombinski@uckwum.pl, Department of Pediatric Radiology, Medical University of Warsaw, 63A Żwirki i Wigury, Warsaw 02-091, Poland}}}
\author[2]{\small \textbf{Patryk Szatkowski}}
\author[3, 4]{\small \textbf{Bartłomiej Sobieski}}
\author[3, 4]{\small \textbf{Tymoteusz Kwieciński}}
\author[3, 4, 5, 6]{\small \textbf{Szymon Płotka}}
\author[7, 8]{\small \textbf{Mariusz Adamek}}
\author[9]{\small \textbf{Marcin Banasiuk}}
\author[1]{\small \textbf{Mariusz I. Furmanek}}
\author[3, 4, 10]{\small \textbf{Przemysław Biecek}}

\affil[1]{Department of Pediatric Radiology, Medical University of Warsaw, Poland}
\affil[2]{2nd Department of Clinical Radiology, Medical University of Warsaw, Poland}
\affil[3]{Faculty of Mathematics and Information Science, Warsaw University of Technology, Poland}
\affil[4]{MI2.ai}
\affil[5]{Informatics Institute, University of Amsterdam, The Netherlands}
\affil[6]{Department of Biomedical Engineering and Physics, Amsterdam University Medical Center, The Netherlands}
\affil[7]{Medical University of Silesia, Katowice, Poland}
\affil[8]{Medical University of Gdańsk, Poland}
\affil[9]{Department of Pediatric Gastroenterology and Nutrition, Medical University of Warsaw, Poland}
\affil[10]{Faculty of Mathematics, Informatics, and Mechanics, University of Warsaw, Poland}

\begin{document}

\maketitle

\section{ABSTRACT}

\subsection{Purpose}
This study assessed the underestimation of lung regions on chest X-ray segmentation masks created according to the current state-of-the-art method, by comparison with total lung volume evaluated on computed tomography (CT).  

\subsection{Materials and methods}
This retrospective study included data from non-contrast chest low-dose CT (LDCT) examinations of 55 healthy individuals. Synthetic X-ray images were generated by projecting a 3D CT examination onto a 2D image plane. Two experienced radiologists manually created two types of lung masks: 3D lung masks from CT examinations (serving as ground truth for further calculations) and 2D lung masks from synthetic X-ray images (according to the current state-of-the-art method: following the contours of other anatomical structures). Overlapping and non-overlapping lung regions covered by both types of masks were analyzed. The volume of the overlapping regions was compared with total lung volume, and the volume fractions of non-overlapping lung regions in relation to the total lung volume were calculated. The performance results between the two radiologists were compared.

\subsection{Results}
Significant differences were observed between lung regions covered by CT and synthetic X-ray masks. The mean volume fractions of the lung regions not covered by synthetic X-ray masks for the right lung, the left lung, and both lungs were 22.8\%, 32.9\%, and 27.3\%, respectively, for Radiologist 1 and 22.7\%, 32.9\%, and 27.3\%, respectively, for Radiologist 2. There was excellent spatial agreement between the masks created by the two radiologists.

\subsection{Conclusions}
Lung X-ray masks created according to the current state-of-the-art method significantly underestimate lung regions and do not cover substantial portions of the lungs. 

\subsection*{Keywords} lung segmentation, chest x-ray, lung CT, cxr, lung masks

\subsection*{Summary} 
Lung X-ray masks created by following the contours of the heart, mediastinum, and diaphragm significantly underestimate lung regions and exclude substantial portions of the lungs from further assessment, which may result in numerous clinical errors.

\subsection*{Key points} 
    \begin{itemize}
        \item Lung mask creation lacks well-defined criteria and standardized guidelines, leading to a high degree of subjectivity between annotators.
        \item Mean lung volume fractions of 22.8\%, 32.9\%, and 27.3\% for the right lung, the left lung, and both lungs, respectively, are obscured by other chest structures on chest X-ray images.
        \item The integration of additional imaging modalities providing detailed 3D information, such as CT and MRI, has great potential to create more precise and comprehensive 2D lung masks.
    \end{itemize}
    
\subsection*{Funding information}
Work on this project is financially supported from the INFOSTRATEG-I/0022/2021-00 grant funded by Polish National Centre for Research and Development (NCBiR).

This research was carried out with the support of the Laboratory of Bioinformatics and Computational Genomics and the High Performance Computing Center of the Faculty of Mathematics and Information Science Warsaw University of Technology.

\section{INTRODUCTION}

Chest X-ray (CXR) is among the most frequently performed diagnostic imaging examinations, and it is one of the most commonly analyzed by automated detection algorithms (1). Recent advancements in CXR have primarily been driven by the COVID-19 pandemic, resulting in the creation of large-scale, publicly available CXR datasets (2). X-ray images are easier to analyze than those of other imaging modalities because X-ray produces a single 2D image, unlike computed tomography (CT) or magnetic resonance imaging (MRI), which consists of up to several hundred images in 3D form.

The analysis of CXR examinations addresses several key challenges, including detection, classification, and segmentation (3), and establishing a clear delineation between lung and non-lung regions is crucial (4). The initial step in the pre-processing phase—preceding feature extraction and algorithm analysis—is the localization of the relevant region of interest (1,3,5). Masks are created to cover the relevant portion of the image (2,4). On CXRs, these masks effectively separate the lung region from the background, confining the analysis to a specific area and reducing the noise caused by the presence of adjacent structures.

The dominant approach involves outlining anatomical structures along their contours (1,3). This methodology likely stems from the early adoption of artificial intelligence (AI) into diagnostic imaging, in which edge detection techniques played a key role (6-9). As a result, lung masks are generated by following the contours of the heart, mediastinum, diaphragm, and chest wall. This practice applies to both manual segmentation performed by human experts and automatic lung segmentation, and it is currently the state-of-the-art method.

However, the state-of-the-art technique does not consider the three-dimensionality of chest organs. CXR is a projection of 3D structures onto a 2D plane, and thus the portions of the lungs may be positioned behind or in front of other chest organs (i.e., the heart, mediastinum, aorta, and regions around the domes of the diaphragm). The interpretation and segmentation of the CXR require correct reference to spatial relationships of lung anatomy.

Most 2D lung masks presented in existing publications and commercially available AI-based software are incomplete and fail to encompass the whole range of the lungs. This inadequacy has significant implications for lung assessment because it restricts the field of analysis and can potentially result in erroneous conclusions. Numerous lesions and abnormalities may be concealed on CXR behind the heart, the mediastinum, or the domes of the diaphragm (10-11). These regions are referred to as obscured lung regions (12) throughout this study.

A comprehensive 3D perspective of the chest organs is achievable through a CT scan. CT images are analyzed in the transverse plane as well as in reconstructions in the coronal and sagittal planes. The coronal plane aligns with the plane of the X-ray image, serving as an equivalent representation. Projecting a 3D CT examination onto a 2D image plane allows the generation of synthetic X-ray images (i.e., digitally reconstructed radiographs (DRRs)) (13).

This study assessed the underestimation of lung regions on chest X-ray segmentation masks created according to the current state-of-the-art method, by calculating the volume fractions of the obscured lung regions in relation to the total lung volume evaluated on CT. 

\section{MATERIALS AND METHODS}

\subsection{Dataset and sample selection}
This retrospective study was approved by the institutional review board, and informed consent was not required. The study comprised an analysis of non-contrast chest low-dose CT (LDCT) examinations from a national screening program, which was conducted in 60 medical centers between 2010 and 2018. The dataset contains 37,839 anonymized chest CT examinations with original radiology reports and a wide range of metadata (i.e., demographics, smoking history, exposure to toxic substances, and family history of lung diseases). The dataset is still being processed and analyzed, with plans to make the whole database publicly available. Before use, the dataset was anonymized following the ethical standards of the Helsinki Declaration. 

The following inclusion and exclusion criteria were applied:

Inclusion criteria:
    \begin{enumerate}
        \item  Age between 50 and 60 years with no history of lung disease affecting the lung volume (i.e., asthma or emphysema) and no history of smoking.

        \item Non-contrast chest LDCT examinations performed on the same CT scanner with the same scan protocol (i.e., kVp, slice thickness) in a period not longer than 6 months.

        \item No major findings on the original radiological report; the definition of a major finding was any finding that could have influenced the lung volume (i.e., emphysema, atelectasis, or consolidations) or other disease that could have affected the correct proportions of the chest organs and thus the measurements of the lung mask (i.e., cardiomegaly, pleural fluid, pneumothorax, or mediastinal enlargement).
    \end{enumerate}

Exclusion criteria:
    \begin{enumerate}
        \item Any major finding directly verified on the CT examination by the two radiologists participating in the study.

        \item Poor examination quality (e.g., part of the lung was outside the scope of the examination, or severe motion artifacts were present).
    \end{enumerate}
    
A pilot study with one CT examination was performed. According to the results, the sample size was estimated at a minimum of 30 CT examinations.

The database was analyzed, and a homogeneous group of 59 CT examinations that met the inclusion criteria was identified; 4 of them were excluded, and 55 were included in the final analysis (Fig 1).

All analyzed CT examinations were performed on an Optima CT520 Series scanner (GE Medical Systems, Waukesha, Wisconsin, USA) with 120 kVp and a 1.25mm slice thickness. The matrix size in the transverse plane (x–y axes) was 512 × 512. The field of view and scan length were adjusted to the patient size.


\begin{figure}[h!]
    \centering
    \includegraphics[width=0.5\linewidth]{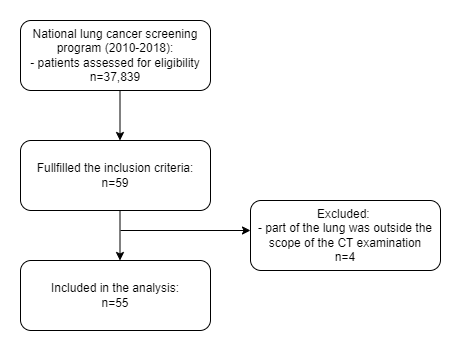}
    \caption{Sample selection flowchart.}
    \label{fig:Figure 1}
\end{figure}

\noindent The following steps in images and lung masks processing were applied (Fig 2):

\subsection*{Step 1: DRR image preparation}
DRR images were generated for each CT examination by averaging the voxel values along the y-axis. The matrix size of the DRR image was consistent with the x–z dimensions of the corresponding CT examination. As a result, pairs of CT/DRR images were obtained.

\subsection*{Step 2: Segmentation performed by radiologists}
Two European Diploma in Radiology (EDiR)-certified radiologists were involved in the study: PBo (Radiologist 1, graduated 8 years prior) and PS (Radiologist 2, newly graduated).    

For each pair of CT and DRR images, the following masks were created with the Slicer 5.2.2 platform (14):
    \begin{itemize}
        \item A 3D CT mask: A semi-automatic segmentation module (Chest Imaging Platform—Lung CT Segmenter) was used to generate 3D lung masks from CT examinations. These masks served as ground truth for further calculations. 

        \item A 2D DRR mask: The Segment Editor module was used to manually create 2D lung masks from DRR according to the current state-of-the-art method.
    \end{itemize}
    
\noindent The radiologists created each type of mask independently.  

\subsection*{Step 3: Mask array transformations}
Python 3.9.12 programming language was used for the following transformations and calculations made on mask arrays.

First, 2D DRR masks were reshaped to the 3D form by multiplying 512 times in the y dimension. The resulting 3D DRR mask sizes corresponded exactly to the 3D CT masks.

Second, 3D CT masks were reshaped to the 2D form by projecting them onto the coronal plane. The resulting 2D CT mask sizes corresponded exactly to the 2D DRR masks.

The final set consisted of four types of masks for each pair of CT/DRR images.

\subsection*{Step 4: 3D mask subtraction}
To compare 3D CT masks with 3D DRR masks, overlapping and non-overlapping regions were calculated. The overlapping regions corresponded to lung regions covered by masks generated from the DRR images. Non-overlapping regions within the 3D CT masks corresponded to obscured lung regions. The 3D CT masks, which served as the ground truth, were used to calculate the total lung volume.

\FloatBarrier
\begin{figure}[h!]
    \centering
    \includegraphics[width=0.8\linewidth]{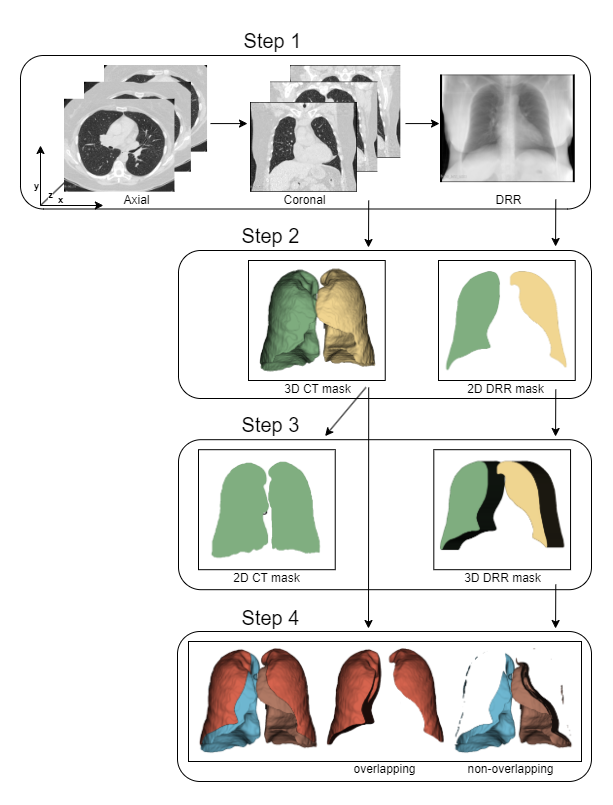}
    \caption{Diagram of subsequent steps in images and lung masks processing.}
    \label{fig:Figure 2}
\end{figure}
\FloatBarrier

\subsection{Statistical analysis}
Statistical analyses were performed in Statistica 13.3 (Tibco Soft Inc, Tulsa, Oklahoma, USA).

The Dice Score Coefficient (DSC) and Jaccard Index (JI) were used to estimate the spatial agreement between the masks created by the two radiologists. The DSC was calculated as follows: 2 times the number of overlapping pixels divided by the sum of pixels in both images. The JI was calculated as follows: the number of overlapping pixels divided by the sum of overlapping and non-overlapping pixels in both images.

To compare the lung regions covered by the CT and DRR masks, lung volume was calculated on the basis of the voxel dimensions. First, the pixel spacing (specified by two numeric values of pixel dimensions in millimeters) was collected from the DICOM metadata (Table 1). Second, the pixel area in the transverse plane was calculated by multiplying the pixel spacing dimensions. Third, voxel volume was calculated by multiplying the pixel area by the slice thickness, which was 1.25 mm in all CT exams. The results are presented in milliliters. The volume calculated from the 3D CT mask corresponded to the total lung volume. The volume of the overlapping regions corresponded to the lung regions covered by the DRR masks. The volume of the non-overlapping regions within the 3D CT masks corresponded to the obscured lung regions; their volume fractions were calculated as follows: the volume of the obscured lung region was divided by total lung volume. This was calculated separately for both lungs and both radiologists. The results are presented as percentages.

The Shapiro–Wilk and Kolmogorov–Smirnov tests were used to test the distributions of the continuous variables. Dependent samples t-tests were used to compare volume and fractions. A Wilcoxon test was used, if appropriate. Differences with a p-value of \textless0.05 were considered statistically significant.

\section{RESULTS}
A total of 55 patients (mean age 55.5, range 50–60; 20 male, 35 female) were included in the study (Table 1).

The masks created by the two radiologists exhibited excellent special agreement (Table 2). For the 3D CT masks, the median DSC scores and median JI scores were 0.999 for all measurements. For the 2D DRR masks, the median DSC scores and median JI scores were 0.973 and 0.948, 0.968 and 0.937, and 0.972 and 0.945 for the right lung, the left lung, and both lungs, respectively.

The results for lung volume calculated from the CT and DRR masks are presented for the right lung, the left lung, and both lungs and for both radiologists (Table 3). In all cases, the total lung volume calculated from CT masks and the lung volume calculated from DRR masks differed significantly. 

The results for the volume fractions of the obscured lung regions are presented separately for the right lung, the left lung, and both lungs and for both radiologists (Table 4). The mean fraction was greater for the left lung (32.9\%, the same for both radiologists) than the right lung (22.8\% and 22.7\% for Radiologist 1 and Radiologist 2, respectively). The mean fraction for both lungs was 27.3\% and was the same for both radiologists. No significant differences were observed between radiologists for any of the comparisons.

\FloatBarrier 
\begin{table}[h] 
\centering
\begin{adjustbox}{width=0.4\textwidth,center=\textwidth}
\small
\begin{tabular}{|l|c|c|c|c|}
\hline
                   & \multicolumn{1}{l|}{\textbf{mean}} & \multicolumn{1}{l|}{\textbf{SD}} & \multicolumn{1}{l|}{\textbf{min}} & \multicolumn{1}{l|}{\textbf{max}} \\ \hline
\textbf{Patients:} &                                    &                                  &                                   &                                   \\ \hline
age                & 55.5                               & 2,8                              & 50                                & 60                                \\ \hline
\textbf{CT exams:} &                                    &                                  &                                   &                                   \\ \hline
pixel spacing *    & 0.66                               & 0.06                             & 0.52                              & 0.80                              \\ \hline
number of slices   & 244                                & 22                               & 195                               & 283                               \\ \hline
scan length **     & 305                                & 28                               & 244                               & 354                               \\ \hline
\end{tabular}
\end{adjustbox}
\caption{Patients and CT exams statistics; * pixel spacing in millimeters in the transverse plane, with one dimension provided; pixels in all analyzed CT exams were square; ** scan length in millimeters was calculated as follows: number of slices multiplied by slice thickness, which was 1.25 mm in all analyzed CT exams.}
\label{tab:Table 1}
\end{table}
\FloatBarrier

\FloatBarrier
\begin{table}[h!]
\centering
\begin{adjustbox}{width=0.8\textwidth,center=\textwidth}
\begin{tabular}{lcccccccccc}
\cline{2-11}
\multicolumn{1}{l|}{}                     & \multicolumn{10}{c|}{\textbf{3D CT masks}}                                                                                                                                                                                                                                                                                                                               \\ \cline{2-11} 
\multicolumn{1}{l|}{}                     & \multicolumn{5}{c|}{\textbf{DSC}}                                                                                                                                                  & \multicolumn{5}{c|}{\textbf{JI}}                                                                                                                                                   \\ \cline{2-11} 
\multicolumn{1}{l|}{}                     & \multicolumn{1}{c|}{\textbf{median}} & \multicolumn{1}{c|}{\textbf{Q1}} & \multicolumn{1}{c|}{\textbf{Q3}} & \multicolumn{1}{c|}{\textbf{min}} & \multicolumn{1}{c|}{\textbf{max}} & \multicolumn{1}{c|}{\textbf{median}} & \multicolumn{1}{c|}{\textbf{Q1}} & \multicolumn{1}{c|}{\textbf{Q3}} & \multicolumn{1}{c|}{\textbf{min}} & \multicolumn{1}{c|}{\textbf{max}} \\ \hline
\multicolumn{1}{|l|}{\textbf{Right lung}} & \multicolumn{1}{c|}{0.999}           & \multicolumn{1}{c|}{0.999}       & \multicolumn{1}{c|}{1.000}       & \multicolumn{1}{c|}{0.990}        & \multicolumn{1}{c|}{1.000}        & \multicolumn{1}{c|}{0.999}           & \multicolumn{1}{c|}{0.998}       & \multicolumn{1}{c|}{0.999}       & \multicolumn{1}{c|}{0.980}        & \multicolumn{1}{c|}{1.000}        \\ \hline
\multicolumn{1}{|l|}{\textbf{Left lung}}  & \multicolumn{1}{c|}{0.999}           & \multicolumn{1}{c|}{0.999}       & \multicolumn{1}{c|}{1.000}       & \multicolumn{1}{c|}{0.992}        & \multicolumn{1}{c|}{1.000}        & \multicolumn{1}{c|}{0.999}           & \multicolumn{1}{c|}{0.998}       & \multicolumn{1}{c|}{1.000}       & \multicolumn{1}{c|}{0.985}        & \multicolumn{1}{c|}{1.000}        \\ \hline
\multicolumn{1}{|l|}{\textbf{Both lungs}} & \multicolumn{1}{c|}{0.999}           & \multicolumn{1}{c|}{0.999}       & \multicolumn{1}{c|}{1.000}       & \multicolumn{1}{c|}{0.993}        & \multicolumn{1}{c|}{1.000}        & \multicolumn{1}{c|}{0.999}           & \multicolumn{1}{c|}{0.998}       & \multicolumn{1}{c|}{0.999}       & \multicolumn{1}{c|}{0.986}        & \multicolumn{1}{c|}{1.000}        \\ \hline
                                          & \multicolumn{1}{l}{}                 & \multicolumn{1}{l}{}             & \multicolumn{1}{l}{}             & \multicolumn{1}{l}{}              & \multicolumn{1}{l}{}              & \multicolumn{1}{l}{}                 & \multicolumn{1}{l}{}             & \multicolumn{1}{l}{}             & \multicolumn{1}{l}{}              & \multicolumn{1}{l}{}              \\ \cline{2-11} 
\multicolumn{1}{l|}{}                     & \multicolumn{10}{c|}{\textbf{2D DRR masks}}                                                                                                                                                                                                                                                                                                                              \\ \cline{2-11} 
\multicolumn{1}{l|}{}                     & \multicolumn{5}{c|}{\textbf{DSC}}                                                                                                                                                  & \multicolumn{5}{c|}{\textbf{JI}}                                                                                                                                                   \\ \cline{2-11} 
\multicolumn{1}{l|}{}                     & \multicolumn{1}{c|}{\textbf{median}} & \multicolumn{1}{c|}{\textbf{Q1}} & \multicolumn{1}{c|}{\textbf{Q3}} & \multicolumn{1}{c|}{\textbf{min}} & \multicolumn{1}{c|}{\textbf{max}} & \multicolumn{1}{c|}{\textbf{median}} & \multicolumn{1}{c|}{\textbf{Q1}} & \multicolumn{1}{c|}{\textbf{Q3}} & \multicolumn{1}{c|}{\textbf{min}} & \multicolumn{1}{c|}{\textbf{max}} \\ \hline
\multicolumn{1}{|l|}{\textbf{Right lung}} & \multicolumn{1}{l|}{0.973}           & \multicolumn{1}{l|}{0.964}       & \multicolumn{1}{l|}{0.980}       & \multicolumn{1}{l|}{0.923}        & \multicolumn{1}{l|}{0.987}        & \multicolumn{1}{l|}{0.948}           & \multicolumn{1}{l|}{0.930}       & \multicolumn{1}{l|}{0.960}       & \multicolumn{1}{l|}{0.858}        & \multicolumn{1}{l|}{0.975}        \\ \hline
\multicolumn{1}{|l|}{\textbf{Left lung}}  & \multicolumn{1}{l|}{0.968}           & \multicolumn{1}{l|}{0.960}       & \multicolumn{1}{l|}{0.973}       & \multicolumn{1}{l|}{0.917}        & \multicolumn{1}{l|}{0.983}        & \multicolumn{1}{l|}{0.937}           & \multicolumn{1}{l|}{0.923}       & \multicolumn{1}{l|}{0.948}       & \multicolumn{1}{l|}{0.846}        & \multicolumn{1}{l|}{0.967}        \\ \hline
\multicolumn{1}{|l|}{\textbf{Both lungs}} & \multicolumn{1}{l|}{0.972}           & \multicolumn{1}{l|}{0.964}       & \multicolumn{1}{l|}{0.975}       & \multicolumn{1}{l|}{0.929}        & \multicolumn{1}{l|}{0.982}        & \multicolumn{1}{l|}{0.945}           & \multicolumn{1}{l|}{0.931}       & \multicolumn{1}{l|}{0.952}       & \multicolumn{1}{l|}{0.868}        & \multicolumn{1}{l|}{0.965}        \\ \hline
\end{tabular}
\end{adjustbox}
\caption{Evaluation of spatial agreement between masks created by both radiologists. Results presented are the Dice Score Coefficient (DSC) and Jaccard Index (JI) for the right lung, left lung, and both lungs and for both types of masks. }
\label{tab:Table 2}
\end{table}
\FloatBarrier


\begin{table}[h!]
\centering
\begin{adjustbox}{width=0.8\textwidth,center=\textwidth}
\begin{tabular}{lcccccccc}
\cline{2-9}
\multicolumn{1}{l|}{}                     & \multicolumn{8}{c|}{\textbf{Radiologist 1}}                                                                                                                                                                                                                                                   \\ \cline{2-9} 
\multicolumn{1}{l|}{}                     & \multicolumn{4}{c|}{\textbf{CT mask volume {[}ml{]}}}                                                                                         & \multicolumn{4}{c|}{\textbf{DRR mask volume {[}ml{]}}}                                                                                        \\ \cline{2-9} 
\multicolumn{1}{l|}{}                     & \multicolumn{1}{c|}{\textbf{mean}} & \multicolumn{1}{c|}{\textbf{SD}} & \multicolumn{1}{c|}{\textbf{min}} & \multicolumn{1}{c|}{\textbf{max}} & \multicolumn{1}{c|}{\textbf{mean}} & \multicolumn{1}{c|}{\textbf{SD}} & \multicolumn{1}{c|}{\textbf{min}} & \multicolumn{1}{c|}{\textbf{max}} \\ \hline
\multicolumn{1}{|l|}{\textbf{Right lung}} & \multicolumn{1}{c|}{2662.0}        & \multicolumn{1}{c|}{610.4}       & \multicolumn{1}{c|}{1466.2}       & \multicolumn{1}{c|}{4017.9}       & \multicolumn{1}{c|}{2058.3}        & \multicolumn{1}{c|}{497.2}       & \multicolumn{1}{c|}{1205.0}       & \multicolumn{1}{c|}{3186.2}       \\ \hline
\multicolumn{1}{|l|}{\textbf{Left lung}}  & \multicolumn{1}{c|}{2275.7}        & \multicolumn{1}{c|}{587.3}       & \multicolumn{1}{c|}{1233.6}       & \multicolumn{1}{c|}{3845.9}       & \multicolumn{1}{c|}{1540.3}        & \multicolumn{1}{c|}{486.1}       & \multicolumn{1}{c|}{616.2}        & \multicolumn{1}{c|}{3120.8}       \\ \hline
\multicolumn{1}{|l|}{\textbf{Both lungs}} & \multicolumn{1}{c|}{4937.7}        & \multicolumn{1}{c|}{1188.0}      & \multicolumn{1}{c|}{2699.7}       & \multicolumn{1}{c|}{7803.8}       & \multicolumn{1}{c|}{3598.6}        & \multicolumn{1}{c|}{956.8}       & \multicolumn{1}{c|}{1821.2}       & \multicolumn{1}{c|}{6306.9}       \\ \hline
                                          & \multicolumn{1}{l}{}               & \multicolumn{1}{l}{}             & \multicolumn{1}{l}{}              & \multicolumn{1}{l}{}              & \multicolumn{1}{l}{}               & \multicolumn{1}{l}{}             & \multicolumn{1}{l}{}              & \multicolumn{1}{l}{}              \\ \cline{2-9} 
\multicolumn{1}{l|}{}                     & \multicolumn{8}{c|}{\textbf{Radiologist 2}}                                                                                                                                                                                                                                                   \\ \cline{2-9} 
\multicolumn{1}{l|}{}                     & \multicolumn{4}{c|}{\textbf{CT mask volume {[}ml{]}}}                                                                                         & \multicolumn{4}{c|}{\textbf{DRR mask volume {[}ml{]}}}                                                                                        \\ \cline{2-9} 
\multicolumn{1}{l|}{}                     & \multicolumn{1}{c|}{\textbf{mean}} & \multicolumn{1}{c|}{\textbf{SD}} & \multicolumn{1}{c|}{\textbf{min}} & \multicolumn{1}{c|}{\textbf{max}} & \multicolumn{1}{c|}{\textbf{mean}} & \multicolumn{1}{c|}{\textbf{SD}} & \multicolumn{1}{c|}{\textbf{min}} & \multicolumn{1}{c|}{\textbf{max}} \\ \hline
\multicolumn{1}{|l|}{\textbf{Right lung}} & \multicolumn{1}{c|}{2660.7}        & \multicolumn{1}{c|}{611.0}       & \multicolumn{1}{c|}{1466.0}       & \multicolumn{1}{c|}{4018.7}       & \multicolumn{1}{c|}{2066.4}        & \multicolumn{1}{c|}{520.1}       & \multicolumn{1}{c|}{1177.9}       & \multicolumn{1}{c|}{3237.9}       \\ \hline
\multicolumn{1}{|l|}{\textbf{Left lung}}  & \multicolumn{1}{c|}{2275.5}        & \multicolumn{1}{c|}{587.1}       & \multicolumn{1}{c|}{1234.0}       & \multicolumn{1}{c|}{3848.2}       & \multicolumn{1}{c|}{1539.2}        & \multicolumn{1}{c|}{488.4}       & \multicolumn{1}{c|}{621.4}        & \multicolumn{1}{c|}{3187.1}       \\ \hline
\multicolumn{1}{|l|}{\textbf{Both lungs}} & \multicolumn{1}{c|}{4936.1}        & \multicolumn{1}{c|}{1188.5}      & \multicolumn{1}{c|}{2700.1}       & \multicolumn{1}{c|}{7807.9}       & \multicolumn{1}{c|}{3605.6}        & \multicolumn{1}{c|}{979.8}       & \multicolumn{1}{c|}{1807.4}       & \multicolumn{1}{c|}{6425.0}       \\ \hline
\end{tabular}
\end{adjustbox}
\caption{Evaluation of lung volume calculated from CT and DRR masks. CT masks, serving as ground truth, were used to calculate total lung volume. Results are presented in milliliters for the right lung, left lung, and both lungs and for both radiologists. In all cases, the total lung volume calculated from CT masks and the lung volume calculated from DRR masks differed significantly.}
\label{tab:Table 3}
\end{table}


\begin{table}[h!]
\centering
\begin{adjustbox}{width=0.8\textwidth,center=\textwidth}
\begin{tabular}{l|llll|llll|l}
\cline{2-9}
                                          & \multicolumn{4}{c|}{\textbf{Radiologist 1}}                                                                                                   & \multicolumn{4}{c|}{\textbf{Radiologist   2}}                                                                                                 & \multicolumn{1}{c}{\textbf{}}         \\ \cline{2-10} 
                                          & \multicolumn{1}{c|}{\textbf{mean}} & \multicolumn{1}{c|}{\textbf{SD}} & \multicolumn{1}{c|}{\textbf{min}} & \multicolumn{1}{c|}{\textbf{max}} & \multicolumn{1}{c|}{\textbf{mean}} & \multicolumn{1}{c|}{\textbf{SD}} & \multicolumn{1}{c|}{\textbf{min}} & \multicolumn{1}{c|}{\textbf{max}} & \multicolumn{1}{c|}{\textbf{p value}} \\ \hline
\multicolumn{1}{|l|}{\textbf{Right lung}} & \multicolumn{1}{l|}{22.8}          & \multicolumn{1}{l|}{4.9}         & \multicolumn{1}{l|}{11.1}         & 32.6                              & \multicolumn{1}{l|}{22.7}          & \multicolumn{1}{l|}{4.8}         & \multicolumn{1}{l|}{13.8}         & 30.7                              & \multicolumn{1}{l|}{0.679}            \\ \hline
\multicolumn{1}{|l|}{\textbf{Left lung}}  & \multicolumn{1}{l|}{32.9}          & \multicolumn{1}{l|}{7.3}         & \multicolumn{1}{l|}{18.9}         & 55.9                              & \multicolumn{1}{l|}{32.9}          & \multicolumn{1}{l|}{7.4}         & \multicolumn{1}{l|}{17.2}         & 53.6                              & \multicolumn{1}{l|}{0.678}            \\ \hline
\multicolumn{1}{|l|}{\textbf{Both lungs}} & \multicolumn{1}{l|}{27.3}          & \multicolumn{1}{l|}{5.1}         & \multicolumn{1}{l|}{19.1}         & 40.9                              & \multicolumn{1}{l|}{27.3}          & \multicolumn{1}{l|}{5.1}         & \multicolumn{1}{l|}{17.7}         & 39.0                              & \multicolumn{1}{l|}{0.862}            \\ \hline
\end{tabular}
\end{adjustbox}
\caption{Evaluation of volume fractions of the obscured lung regions in relation to the total lung volume calculated from CT masks. The results are presented as percentages {[}\%{]} for the right lung, left lung, and both lungs and for both radiologists. }
\label{tab: Table 4}
\end{table}

\section{DISCUSSION}
Our study demonstrated that lung X-ray masks created according to the current state-of-the-art method significantly underestimate lung regions and do not cover substantial portions of the lungs. The differences were 22.7\%–22.8\%, 32.9\% and 27,3\% for the right lung, the left lung, and both lungs, respectively.

A previous study conducted in the 1990s estimated the lung volume obscured by other chest structures (12), and the results were similar to ours (with a mean of 26.4\% of the lung volume obscured). The role of proper mask creation has been extensively analyzed in previous publications (1,3,5), although these analyses have primarily focused on image analysis aspects, such as edge detection or pixel-based techniques, and not on anatomical correctness. From a radiological perspective, underestimation can result in numerous clinical errors because significant areas lie outside the masks (10-11). 

Although a comprehensive analysis of available CXR databases is beyond the scope of this publication, a common trend is observed in the literature—lung masks are created along the contours of adjacent organs. This pattern applies to both manual segmentation carried out by radiologists and automatic segmentation, which heavily rely on manual masks as a ground truth (5,15-16). This practice may be driven by the assumption that contour-based lung masks offer a reasonable approximation of the lung boundaries.

The manual segmentation of the lung fields, especially when performed by experienced radiologists, is highly valuable and accurate (5,15-16), although it is relatively uncommon because of the requirement for expertise, time, and increased costs associated with the manual labeling process. Consequently, it is challenging to apply manual labeling to a large number of CXRs (4). By contrast, automatic segmentation techniques offer faster processing and can handle large datasets, but they are susceptible to errors (4). Various techniques have been implemented; these methods commonly utilize conventional feature extraction methods such as edge detection, pixel intensity, and geometric shape analysis, as well as automatic feature extraction with deep learning models (1-5,15,17). Notably, manual masks serve as the reference and ground truth for the learning process and evaluating the performance of automatic segmentation methods. Thus, any ambiguities or errors introduced by human annotators subsequently propagate to later automated solutions.

However, mask creation lacks well-defined criteria and standardized guidelines (18), leading to a high degree of subjectivity between annotators (1). Consequently, lung masks in the same publications and databases are often marked along the contours of different anatomic structures. The most striking example is evident in the area behind the heart. Although most masks are created along the contours of the heart, masks can also be generated along the contours of the aorta or vertebral column, both of which are located behind the heart. Therefore, a mix of masks created along the contours of the heart, aorta, and vertebral column can be observed, which confirms the variability in the segmentation approach. Moreover, the reviewed publications and databases have exceptionally assessed the regions around the domes of the diaphragm. 

Our results have significant implications for clinical practice. First, standard lung masks fail to encompass the whole range of the lungs and significantly restrict the field of analysis. The latter may lead to false conclusions and diagnoses (10-11). This has been the subject of several publications, which have demonstrated the importance of extracting the whole lung region for effective disease detection (19). Second, it is necessary to discuss and review the available datasets to correct and categorize masks on the basis of the specific anatomical structures from which they were prepared. Lung masks generated with different reference boundaries, such as those based on the contours of the heart, aorta, or vertebral column, may cover distinct areas of the lungs. As a result, it is questionable whether these masks can be collectively analyzed without accounting for their variations. Finally, despite the increasing number of publications highlighting the application of AI in diagnostic imaging, a more critical perspective must be adopted. Alongside the hype surrounding AI is growing evidence identifying challenges of its implementation (20-21). This also warrants more active involvement of radiologists, who not only perform the assigned tasks but also critically analyze the methodology and results.

To address the considerations of how to proceed with the CXR lung masks, we propose three potential solutions:
    \begin{enumerate}
        \item Creating one common mask: a single unified mask that includes both the obscured lung regions and the regions covered by the currently generated standard masks (Fig 3C). This process would result in a comprehensive mask representing the whole lung anatomy, including the visible and obscured regions.

        \item Creating two separate masks: one covering the range of lungs currently represented by the standard masks, and the other specifically dedicated to the obscured lung regions (Fig 3B,D). This approach would allow the targeted analysis to be conducted independently, providing valuable insights into any abnormalities present in these obscured lung regions. It would also allow the combination of the currently generated standard lung masks with additional annotations.

        \item Determining the standardized chest region encompassing the lungs. This region may be limited by the rib cage, although it would be challenging to establish the standard lower border of the lungs because of variations in breathing phases, as well as variability in gender or age. Nonetheless, it presents an interesting solution when dealing with CXRs containing complex lung abnormalities (16,22). 
    \end{enumerate}

\FloatBarrier
\begin{figure}[h!]
    \centering
    \includegraphics[width=1\linewidth]{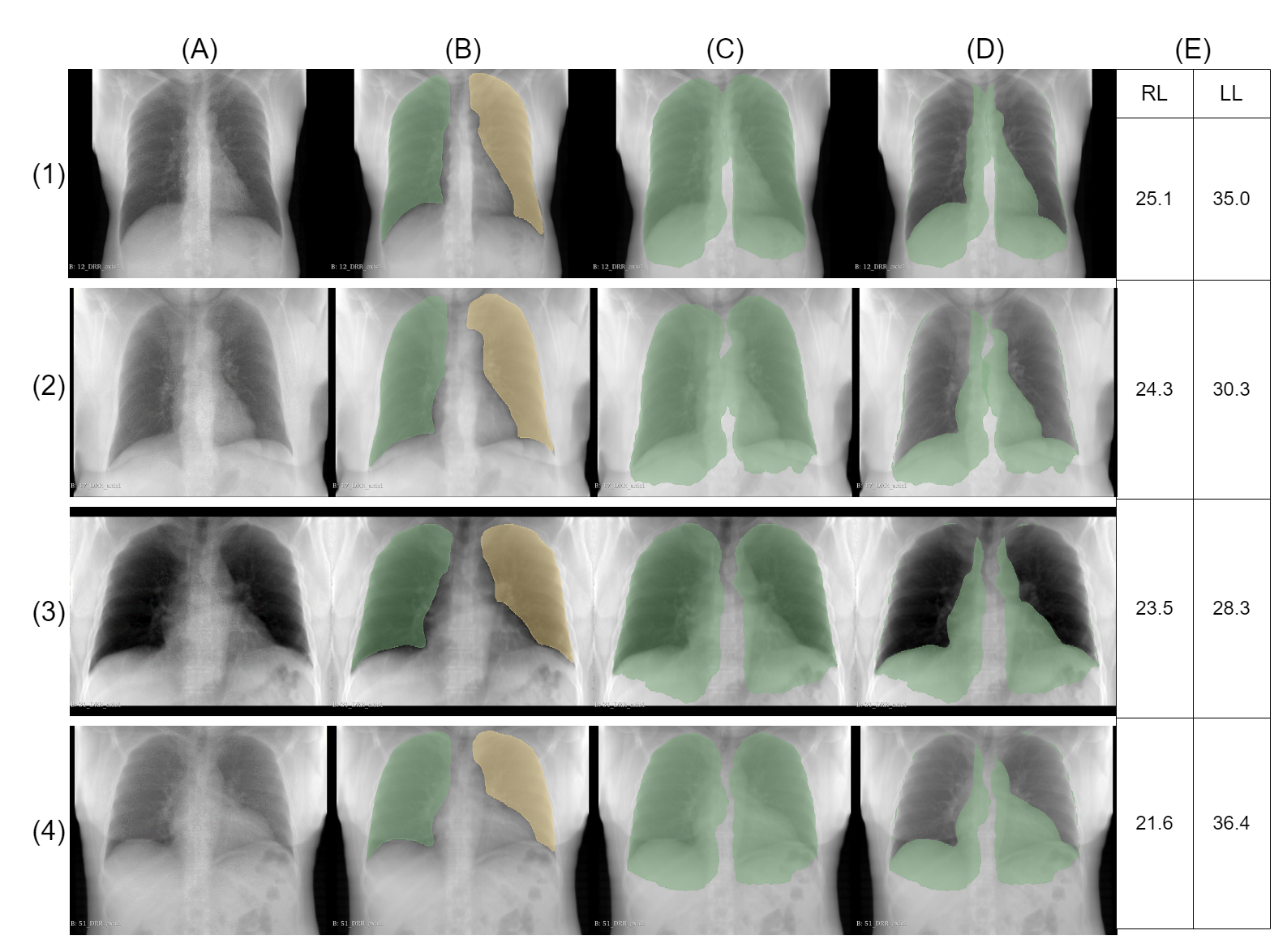} 
    \caption{Sample images from two females (1–2) and two males (3–4): (A) synthetic X-ray—DRR image; (B) 2D DRR mask manually generated by Radiologist 1; (C) 2D CT mask generated by projecting a 3D CT examination onto a 2D image plane; (D) mismatch between previous masks corresponding to obscured lung regions; (E) volume fraction of the obscured lung region in relation to the total lung volume. Results are presented as percentages; RL = right lung, LL = left lung.}
    \label{fig:Figure 3}
\end{figure}
\FloatBarrier

Our study has several limitations:
    \begin{enumerate}
        \item DRRs were used instead of original CXR images. The concept of generating DRR images dates back to the early 1990s, with various algorithms proposed for different cases (13,23-26). Ren (26) described a similar approach to ours to address the problem of bone suppression on CXRs. Thus, the methodology employed in our publication is well-established and has been used in previously published studies.

        \item The examination position may affect the relative position of the chest organs. CT examinations are consistently conducted with patients in the supine position, whereas CXR images can be obtained with the patient in either the standing or supine position. Consequently, discrepancies may arise between the DRRs generated from supine CT scans and CXR images obtained in the standing position.

        \item A selected homogeneous group of healthy patients was analyzed. We did not assess examinations of patients with abnormalities such as pneumonia, pleural fluid, cardiomegaly, or other chest pathologies that could have impacted the measurements. In addition, results in other groups of patients may vary because of differences in the shape and size of the lungs and heart between ages, genders (15), body position during the examination, and the presence or absence of chest comorbidities (1).
    \end{enumerate}
    
In conclusion, our study underlines the importance of proper lung segmentation, as a significant portion of the lung volume remaining outside the mask region may prevent reliable diagnosis. This study also addresses several vital steps toward improving the lung segmentation process: first, the development of standardized criteria for mask creation that would minimize subjectivity and variations in the annotation process between annotators; second, increasing the training and expertise of radiologists and annotators involved in manual lung mask creation, which would lead to better quality of ground truth data; third, the development and improvement of automatic lung segmentation algorithms, which could be trained on datasets that include examples of masks covering the whole area of the lungs. The integration of additional imaging modalities providing detailed 3D information, such as CT or MRI, has great potential to create more precise and comprehensive 2D lung masks.

\section{References}

    \begin{enumerate}
        \item Candemir S, Antani S. A review on lung boundary detection in chest X-rays. Int J Comput Assist Radiol Surg. 2019 Apr;14(4):563-576. doi: 10.1007/s11548-019-01917-1. Epub 2019 Feb 7. PMID: 30730032; PMCID: PMC6420899.
        \item Ait Nasser A, Akhloufi MA. A Review of Recent Advances in Deep Learning Models for Chest Disease Detection Using Radiography. Diagnostics (Basel). 2023 Jan 3;13(1):159. doi: 10.3390/diagnostics13010159. PMID: 36611451; PMCID: PMC9818166.
        \item F. Munawar, S. Azmat, T. Iqbal, C. Grönlund and H. Ali, "Segmentation of Lungs in Chest X-Ray Image Using Generative Adversarial Networks," in IEEE Access, vol. 8, pp. 153535-153545, 2020, doi: 10.1109/ACCESS.2020.3017915. 
        \item Agrawal T, Choudhary P. Segmentation and classification on chest radiography: a systematic survey. Vis Comput. 2023;39(3):875-913. doi: 10.1007/s00371-021-02352-7. Epub 2022 Jan 8. PMID: 35035008; PMCID: PMC8741572. 
        \item van Ginneken B, Stegmann MB, Loog M. Segmentation of anatomical structures in chest radiographs using supervised methods: a comparative study on a public database. Med Image Anal. 2006 Feb;10(1):19-40. doi: 10.1016/j.media.2005.02.002. PMID: 15919232. 
        \item Li L, Zheng Y, Kallergi M, Clark RA. Improved method for automatic identification of lung regions on chest radiographs. Acad Radiol. 2001 Jul;8(7):629-38. doi: 10.1016/S1076-6332(03)80688-8. PMID: 11450964.
        \item Xu XW, Doi K. Image feature analysis for computer-aided diagnosis: detection of right and left hemidiaphragm edges and delineation of lung field in chest radiographs. Med Phys. 1996 Sep;23(9):1613-24. doi: 10.1118/1.597738. PMID: 8892259. 
        \item Cheng, D., Goldberg, M.: An algorithm for segmenting chest radiographs. In: Visual Communications and Image Processing’88: Third in a Series, vol. 1001, pp. 261–268. International Society for Optics and Photonics (1988) 
        \item Duryea J, Boone JM. A fully automated algorithm for the segmentation of lung fields on digital chest radiographic images. Med Phys. 1995 Feb;22(2):183-91. doi: 10.1118/1.597539. PMID: 7565349.
        \item Gefter WB, Post BA, Hatabu H. Commonly Missed Findings on Chest Radiographs: Causes and Consequences. Chest. 2023 Mar;163(3):650-661. doi: 10.1016/j.chest.2022.10.039. Epub 2022 Dec 12. PMID: 36521560; PMCID: PMC10154905.
        \item Del Ciello A, Franchi P, Contegiacomo A, Cicchetti G, Bonomo L, Larici AR. Missed lung cancer: when, where, and why? Diagn Interv Radiol. 2017 Mar-Apr;23(2):118-126. doi: 10.5152/dir.2016.16187. PMID: 28206951; PMCID: PMC5338577.
        \item Chotas HG, Ravin CE. Chest radiography: estimated lung volume and projected area obscured by the heart, mediastinum, and diaphragm. Radiology. 1994 Nov;193(2):403-4. doi: 10.1148/radiology.193.2.7972752. PMID: 7972752.
        \item Levine L, Levine M. DRRGenerator: A Three-dimensional Slicer Extension for the Rapid and Easy Development of Digitally Reconstructed Radiographs. J Clin Imaging Sci. 2020 Oct 29;10:69. doi: 10.25259/JCIS\_105\_2020. PMID: 33194311; PMCID: PMC7656050. 
        \item Kikinis, R., Pieper, S.D., Vosburgh, K.G. (2014). 3D Slicer: A Platform for Subject-Specific Image Analysis, Visualization, and Clinical Support. In: Jolesz, F. (eds) Intraoperative Imaging and Image-Guided Therapy. Springer, New York, NY. https://doi.org/10.1007/978-1-4614-7657-3\_19. 
        \item Liu W, Luo J, Yang Y, Wang W, Deng J, Yu L. Automatic lung segmentation in chest X-ray images using improved U-Net. Sci Rep. 2022 May 23;12(1):8649. doi: 10.1038/s41598-022-12743-y. PMID: 35606509; PMCID: PMC9127108. 
        \item Selvan R, Dam EB, Rischel S, Sheng K, Nielsen M, Pai A. Lung Segmentation from Chest X-rays using Variational Data Imputation. arXiv:2005.10052v2 https://doi.org/10.48550/arXiv.2005.10052. Published v2. July 7, 2020. Accessed January 4, 2024.
        \item Indeewara W, Hennayake M, Rathnayake K, Ambegoda T, Meedeniya D (2023). Chest X-ray Dataset with Lung Segmentation (version 1.0.0). PhysioNet. https://doi.org/10.13026/9cy4-f535. 
        \item Karargyris A, Siegelman J, Tzortzis D, Jaeger S, Candemir S, Xue Z, Santosh KC, Vajda S, Antani S, Folio L, Thoma GR. Combination of texture and shape features to detect pulmonary abnormalities in digital chest X-rays. Int J Comput Assist Radiol Surg. 2016 Jan;11(1):99-106. doi: 10.1007/s11548-015-1242-x. Epub 2015 Jun 20. PMID: 26092662.
        \item Furutani, K., Hirano, Y., and Kido, S., “Segmentation of lung region from chest x-ray images using U-net”, in <i>International Forum on Medical Imaging in Asia 2019</i>, 2019, vol. 11050. doi:10.1117/12.2521594. 
        \item Dratsch T, Chen X, Rezazade Mehrizi M, Kloeckner R, Mähringer-Kunz A, Püsken M, Baeßler B, Sauer S, Maintz D, Pinto Dos Santos D. Automation Bias in Mammography: The Impact of Artificial Intelligence BI-RADS Suggestions on Reader Performance. Radiology. 2023 May;307(4):e222176. doi: 10.1148/radiol.222176. Epub 2023 May 2. PMID: 37129490.
        \item Nazer LH, Zatarah R, Waldrip S, Ke JXC, Moukheiber M, Khanna AK, Hicklen RS, Moukheiber L, Moukheiber D, Ma H, Mathur P. Bias in artificial intelligence algorithms and recommendations for mitigation. PLOS Digit Health. 2023 Jun 22;2(6):e0000278. doi: 10.1371/journal.pdig.0000278. PMID: 37347721; PMCID: PMC10287014.
        \item Reamaroon N, Sjoding MW, Derksen H, Sabeti E, Gryak J, Barbaro RP, Athey BD, Najarian K. Robust segmentation of lung in chest x-ray: applications in analysis of acute respiratory distress syndrome. BMC Med Imaging. 2020 Oct 15;20(1):116. doi: 10.1186/s12880-020-00514-y. PMID: 33059612; PMCID: PMC7566051.
        \item Galvin JM, Sims C, Dominiak G, Cooper JS. The use of digitally reconstructed radiographs for three-dimensional treatment planning and CT-simulation. Int J Radiat Oncol Biol Phys. 1995 Feb 15;31(4):935-42. doi: 10.1016/0360-3016(94)00503-6. PMID: 7860409. 
        \item Moore CS, Avery G, Balcam S, Needler L, Swift A, Beavis AW, Saunderson JR. Use of a digitally reconstructed radiograph-based computer simulation for the optimisation of chest radiographic techniques for computed radiography imaging systems. Br J Radiol. 2012 Sep;85(1017):e630-9. doi: 10.1259/bjr/47377285. Epub 2012 Jan 17. PMID: 22253349; PMCID: PMC3487078.
        \item Milickovic N, Baltast D, Giannouli S, Lahanas M, Zamboglou N. CT imaging based digitally reconstructed radiographs and their application in brachytherapy. Phys Med Biol. 2000 Oct;45(10):2787-800. doi: 10.1088/0031-9155/45/10/305. PMID: 11049172.
        \item Ren G, Xiao H, Lam SK, Yang D, Li T, Teng X, Qin J, Cai J. Deep learning-based bone suppression in chest radiographs using CT-derived features: a feasibility study. Quant Imaging Med Surg. 2021 Dec;11(12):4807-4819. doi: 10.21037/qims-20-1230. PMID: 34888191; PMCID: PMC8611463.
    \end{enumerate}

\end{document}